***Sectional Structure and Emotional Dynamics in Chinese Pop Songs: An Empirical Analysis of Valence–Arousal Trajectories across 100 Songs***

Author: Jingyi Lyu    Affiliation: Communication University of China    Email: ljy@mails.cuc.edu.cn

**Abstract:** Music Emotion Recognition (MER) aims to identify and represent emotional information in music through computational methods and is an important research area within Music Information Retrieval (MIR). To address the limited consideration of song sectional structure in existing dynamic MER research, this study examines 100 Chinese pop songs by aligning 1,046 manually annotated sections with continuous Valence–Arousal (VA) trajectories and analyzing them from the perspectives of section type, adjacent section transitions, repeated sections, and whole-song trajectories. The results show that emotional differences across sections are reflected more strongly in Arousal. Verse typically forms a relatively low-activation baseline, Pre-chorus exhibits a progressive buildup, and Chorus produces a more pronounced high-arousal arrival, while Interlude, Bridge, and Outro tend to show transitional, divergent, and closing functions, respectively. Although whole-song sectional configurations are diverse, high-frequency local transitions are relatively concentrated. High-arousal positions occur more often in the later part of a song, but are not fixed to the final Chorus. Based on these findings, this study summarizes the emotional organization of the selected pop songs as an empirical framework of “local cycles-global accumulation”, in which local sectional cycles are accompanied by emotional pullbacks, while the overall trajectory exhibits a later-stage rise in VA and a tendency for high-Arousal positions to occur later in the song.

**Keywords:** MIR; MER; Valence–Arousal; Sectional Structure; Emotion Trajectories; Chinese Pop Songs

## 1 Introduction

Dynamic music emotion recognition can characterize when and how emotion changes within a song through continuous Valence–Arousal (VA) trajectories, but temporal position alone remains insufficient to explain the functional meaning of these changes within song structure. Musical emotion is not a single label attached to an entire song, but rather a dynamic process that develops, diverges, returns, and resolves over time. The VA model uses Valence to describe emotional polarity and Arousal to describe activation level, and is a commonly used continuous representation of emotion in dynamic MER (Russell, 1980; Eerola & Vuoskoski, 2011). However, similar emotional changes occurring in a Verse, Pre-chorus, or Chorus may correspond to different formal contexts and organizational functions. Therefore, although continuous VA trajectories can indicate when emotional changes occur, temporal position alone cannot explain what these changes mean within the structure of a song.

Research on popular-music form commonly divides songs into functional sections such as Intro, Verse, Chorus, Pre-chorus, Interlude, Bridge, and Outro, with particular emphasis on the relationships among these sections. Summach (2011) argues that Verse typically establishes context and a relatively low-intensity baseline, while Pre-chorus builds momentum through changes in texture, rhythm, register, and harmonic rhythm, thereby strengthening the sense of arrival at the Chorus. de Clercq (2017), meanwhile, emphasizes that formal labels are perceptual and inherently ambiguous and should therefore not be interpreted mechanically outside their specific musical contexts. From this perspective, a gap remains in the existing literature: dynamic MER is well suited to characterizing continuous emotional change but provides limited interpretation of its formal context; research on popular-music form can explain sectional functions, but has rarely used continuous VA trajectories to quantify how these functions correspond to within-song emotional dynamics.

To address this gap, the present study investigates 100 Chinese pop songs by aligning 1,046 manually reviewed functional sections with VA estimates at a 0.5-s frame shift. Emotional variation within songs is examined from the perspectives of section type, adjacent section transitions, repeated sections, and whole-song trajectories. Songs are treated as the primary unit of statistical inference, and singer-clustered bootstrap analyses are used to assess the robustness of the main findings with respect to correlations among multiple songs by the same singer.

The main findings of this study are as follows. First, emotional differences across functional sections are reflected primarily in Arousal: Verse typically forms a relatively low-activation baseline, Pre-chorus shows a progressive buildup, and Chorus forms a more pronounced high-arousal arrival, whereas sectional differences in Valence are less consistent. Second, section types do not correspond to fixed emotional states. VA transition paths into the same section may differ, and repeated occurrences of the same section do not simply return to their initial emotional levels, indicating that formal recurrence does not necessarily imply emotional recurrence. Third, the selected songs exhibit diverse whole-song sectional configurations but relatively concentrated local transitions. Their VA trajectories are not simply monotonic, but show a later-stage rise in Arousal amid local buildups, pullbacks, and renewed arrivals. Finally, high-arousal positions occur more often in the later part of a song but are not fixed to the final Chorus; meanwhile, large VA changes at the transition into Outro and the simultaneous decline in Valence and Arousal from Chorus to Outro suggest that release and closure after high-arousal states also constitute an important part of late-song emotional organization. These findings are further summarized by the empirical framework of “local cycles−global accumulation,” which captures the coexistence of local sectional recurrence and later-stage emotional buildup.

## 2 Background

### 2.1 Continuous VA Modeling and Dynamic Music Emotion Recognition

MER commonly represents musical emotion using either discrete categories or continuous dimensions. Russell's (1980) circumplex model maps emotion onto a two-dimensional space defined by Valence and Arousal, where Valence represents emotional polarity and Arousal represents the level of activation. Compared with discrete emotion categories, continuous dimensional models are better suited to characterizing gradual changes in emotional direction and intensity and can more readily accommodate ambiguous states that fall between prototypical emotion categories (Eerola & Vuoskoski, 2011). Schubert (1999, 2004) applied two-dimensional emotion spaces to continuous measurement and feature modeling of musical responses, while datasets such as 1000 Songs and DEAM further provided time-continuous Valence and Arousal annotations, enabling within-song emotional variation to be represented as dynamic VA trajectories (Soleymani et al., 2013; Aljanaki et al., 2017).

Advances in music representation learning have further improved the capacity of dynamic MER models to capture local changes in audio. MERT learns representations of pitch, timbre, and higher-level musical semantics through large-scale self-supervised pretraining on music audio and has demonstrated strong transferability across a range of music understanding tasks (Li et al., 2024). Previous studies have also segmented continuous audio into local excerpts for feature learning and subsequently modeled sequences of these excerpts for song-level emotion recognition (He & Ferguson, 2022). Such approaches enhance model sensitivity to local temporal variation; however, the excerpts are typically defined by fixed temporal windows and do not directly correspond to functionally meaningful sections such as Verse or Chorus. Dynamic VA trajectories can therefore indicate when emotional changes occur, but temporal position alone does not specify the formal context in which those changes take place.

Resources such as PMEmo further provide dynamic emotion annotations for popular songs and chorus excerpts, offering an additional foundation for continuous emotion prediction in popular music (Zhang et al., 2018). The frozen teacher models used in this study are based on MERT representations and dynamic OpenSMILE features. Their outputs are treated as a consistent measurement source for examining the relationship between VA trajectories and sectional structure.

### 2.2 Formal Functions of Sections in Pop Songs

Intro, Verse, Pre-chorus, Chorus, Bridge, Interlude, and Outro are common functional sections in pop songs. Verse typically serves to develop lyrical narrative and establish context, whereas Chorus often gains recurring structural prominence through lyrical repetition, memorable melodic material, denser texture, or increased vocal intensity. Positioned between Verse and Chorus, Pre-chorus commonly functions to build momentum, delay arrival, and strengthen the goal-directed quality of the subsequent Chorus (Summach, 2011). Nobile (2022) further conceptualizes the Verse–Pre-chorus–Chorus sequence as a directional formal process, in which Pre-chorus propels structural motion and Chorus functions as the corresponding point of arrival. Such formal motion may be realized through harmony, texture, timbre, vocal treatment, and other musical parameters. Bridge often departs from an established rotational pattern in the latter half of a song and creates contextual contrast, while Interlude and Outro may respectively serve functions such as sectional connection, redirection of energy, and terminal closure.

Sectional structure has also begun to be incorporated into emotion modeling in MER. Raboy and Taparugssanagorn (2024) extracted audio and lyrical features from Verse and Chorus sections and combined information from different sections for song-level emotion classification. Their study demonstrates that sectional structure can provide local information that differs from whole-song input. However, the analysis remains centered on a predefined Verse–Chorus–Verse structure and final song-level emotion categories, without further characterizing continuous VA variation across a broader set of functional sections in complete songs or examining how the same section type may vary across repeated occurrences.

## 2.3 Research Gap and Analytical Framework

Based on the research gap outlined above, this study integrates continuous VA trajectories with the functional sectional structure of complete songs, focusing on how different structural positions correspond to continuous emotional changes within songs. Compared with analyses based on fixed temporal windows or a small number of predefined sections, the present study further incorporates multiple functional sections, including Intro, Verse, Pre-chorus, Chorus, Bridge, Interlude, and Outro, while also examining changes in the emotional states of the same section type across repeated occurrences.

Building on the above work, this study first maps continuous VA estimates with a 0.5-s frame shift onto manually annotated section boundaries. Statistical analyses are then conducted from the perspectives of section type, adjacent section transitions, repeated sections, and whole-song trajectories. Section-type analysis examines the overall positions of different functional sections in VA space; adjacent-transition analysis characterizes local emotional changes; repeated-section analysis evaluates whether formal recurrence is accompanied by a return to similar emotional states; and whole-song trajectory analysis examines the relationship between local sectional cycles and broader emotional progression across the song. Songs are treated as the primary unit of statistical inference, and singer-clustered bootstrap analyses are used to assess the robustness of the main findings with respect to correlations among multiple songs by the same singer. Figure 1 summarizes the overall analytical workflow.

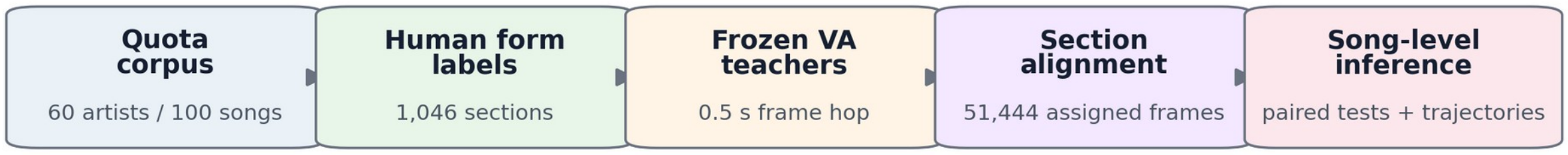

**Figure 1. Section-aware VA analysis workflow for 100 Chinese pop songs.**

# 3 Data and Methods

## 3.1 Corpus and Quota Sampling

The study sample was drawn from the top 60 artists on the QQ Music Singer Chart. A quota sampling scheme was applied by ranking tier: three songs were selected for each artist ranked 1–10, two songs for each artist ranked 11–30, and one song for each artist ranked 31–60, yielding a total of 100 songs. Collaborative tracks were assigned to the relevant artist according to remaining quota availability. The resulting corpus includes 60 quota artists and 68 distinct performer credits, including collaborations. The songs have a mean duration of 257.48 s and contain an average of 10.46 sections. This sampling strategy was designed to represent different chart tiers, while the chart itself provided a general context of market popularity. Specific rankings were used only for quota stratification and were not treated as direct measures of individual-song popularity or artistic quality.

## 3.2 Manual Section Annotation and Quality Control

All 100 Chinese pop songs were manually annotated by the researcher through close listening, with section types and boundaries assigned to seven categories: Intro, Verse, Pre-chorus, Chorus, Bridge, Interlude, and Outro. A total of 1,046 sections were identified. After annotation, 10 songs were reviewed by a reviewer with formal training in music, and the review resulted in no revisions to the assigned section types or boundaries. The final section labels and boundaries were fixed before frame-to-section alignment and statistical analysis. VA inference did not use section boundaries or section labels. Section continuity and frame coverage were also checked, and all 1,046 sections contained valid VA predictions. Because some ambiguity may remain in the interpretation of Bridge, Interlude, and weakly differentiated Verse/Chorus boundaries, the analysis focuses primarily on recurring cross-song tendencies rather than categorical interpretations of individual sections.

## 3.3 Frame-Level VA Inference and Section Alignment

Valence and Arousal were estimated using two audio-only teacher models with identical architectures but trained independently and kept frozen during inference. At each time step, the models concatenate a 768-dimensional representation from the sixth hidden layer of MERT-v1-95M with 260-dimensional dynamic OpenSMILE ComParE_2016 features. The concatenated representation is projected to 112 dimensions and then fed into a global–local BiGRU, producing separate Valence and Arousal outputs in the range of −1 to 1. The teacher models were trained and selected using the PMEmo dataset. On the locked test set, the segment-level PCC/CCC values were 0.8029/0.8002 for Valence and 0.8574/0.8548 for Arousal; at the 0.5-s frame level, the corresponding values were 0.8106/0.8016 and 0.8487/0.8392. None of the 100 songs analyzed in the present study were used for teacher-model training, validation, or testing. Accordingly, the model outputs are treated as model-estimated VA rather than human-annotated continuous emotion ground truth.

For full-song inference, a 30-s sliding window with a 15-s hop was used, and predictions from overlapping windows were fused using Hann-weighted overlap aggregation to generate continuous VA trajectories at a temporal resolution of 0.5 s. Across the 100 songs, 51,448 prediction frames were obtained. Section boundaries, section labels, and human emotion annotations were not used during inference. After inference, the frame-level predictions were mapped to the finalized manual section boundaries. A total of 51,444 frames were successfully assigned to the 1,046 functional sections, while four trailing frames extending beyond the final annotated boundaries were excluded. The Valence and Arousal values of each section were defined as the arithmetic mean of its constituent frame-level predictions and were subsequently used for analyses of section types, adjacent transitions, repeated sections, and whole-song trajectories.

### 3.4 Derived Measures and Statistical Analysis

Section-level analyses used mean Valence, mean Arousal, and changes in VA between adjacent sections. Let the mean VA coordinates of section s in song i be $(V_{i,s+1}, A_{i,s+1})$. Changes between adjacent sections were defined as

$$\Delta V_{i,s} = V_{i,s+1} - V_{i,s}$$

$$\Delta A_{i,s} = A_{i,s+1} - A_{i,s}$$

For consecutive Verse → Pre-chorus → Chorus sequences, the Arousal changes from Verse to Pre-chorus and from Pre-chorus to Chorus were calculated separately to characterize the progression across the two transitions. For transitions into Chorus, $|\Delta V|<0.05$, $|\Delta A|<0.05$were additionally used as descriptive thresholds for “approximate continuity.” These thresholds were used only for pattern counting and were not involved in significance testing.

Whole-song emotional paths were constructed according to section order. For song $i$, the path length $L_i$, displacement between the first and last sections $D_i$, and path directness $R_i$ were defined as

$$L_i = \sum_{s=1}^{S_i - 1} \sqrt{(V_{i,s+1} - V_{i,s})^2 + (A_{i,s+1} - A_{i,s})^2}$$

$$D_i = \sqrt{(V_{i,S_i} - V_{i,1})^2 + (A_{i,S_i} - A_{i,1})^2}, R_i = \frac{D_i}{L_i}$$

Here, $R_i$ describes the directness of the whole-song VA path relative to the displacement between its starting and ending points. For each song, we also recorded the position and section type associated with the highest mean Arousal, whether the final Chorus corresponded to that section, and the largest local change in VA as measured by the Euclidean distance between adjacent sections. Frame-level VA trajectories for complete songs were linearly interpolated to 101 normalized time points for analysis of the average whole-song trajectory.

Songs were treated as the primary unit of statistical inference. For songs containing both section types in a given comparison, within-song paired differences were evaluated using two-sided Wilcoxon signed-rank tests, together with 5,000 nonparametric bootstrap resamples to estimate 95% confidence intervals. Holm correction was applied to multiple comparisons within each comparison family. When the same transition occurred multiple times within a song, transition values were first aggregated within that song before cross-song analysis, thereby reducing weighting differences caused by unequal numbers of repeated transitions. Because some singers contributed multiple songs to the corpus, singer-clustered bootstrap resampling was additionally used as a sensitivity analysis to assess whether the main findings were substantially affected by within-singer dependence.

# 4 Results

## 4.1 Sample Structure: Diverse Whole-Song Forms and Relatively Concentrated Local Transitions

Verse and Chorus occurred in all 100 songs, comprising 246 and 362 sections, respectively. Interlude appeared in 85 songs, whereas Pre-chorus and Bridge appeared in 38 and 43 songs, respectively. Across the 100 songs, 76 distinct raw section sequences were identified. After collapsing consecutive repetitions of the same section type, 49 distinct sequences remained, 36 of which occurred only once. These results indicate that the selected sample does not conform to a single whole-song formal template.

**Table 1. Distribution of Section Types and Descriptive VA Positions**

| Section Type | NO. of sections | No. of Songs | Coverage | Mean V | Mean A | Mean Position |
|---|---|---|---|---|---|---|
| Intro | 95 | 94 | 94% | 0.075 | -0.081 | 0.041 |
| Verse | 246 | 100 | 100% | 0.111 | -0.088 | 0.305 |
| Pre-chorus | 77 | 38 | 38% | 0.187 | 0.107 | 0.381 |
| Chorus | 362 | 100 | 100% | 0.240 | 0.215 | 0.592 |
| Interlude | 121 | 85 | 85% | 0.260 | 0.169 | 0.523 |
| Bridge | 45 | 43 | 43% | 0.273 | 0.270 | 0.689 |
| Outro | 100 | 100 | 100% | 0.114 | 0.104 | 0.953 |

Note: The VA values reported in this table are section-level summaries for descriptive purposes. For subsequent inferential analyses, repeated sections of the same type were first aggregated within each song so that songs, rather than individual sections, determined the statistical weight. Mean position refers to the normalized midpoint of each section.

Compared with the diversity of whole-song section sequences, local section transitions were relatively concentrated. The most frequent transitions were Verse → Chorus (116 occurrences), Chorus → Interlude (111), Chorus → Chorus (100), Intro → Verse (94), Chorus → Outro (91), Pre-chorus → Chorus (76), Interlude → Verse (68), and Verse → Pre-chorus (68). Together, these eight transition types accounted for 76.5% of all 946 adjacent section transitions. These results suggest that, within the selected sample, recurring regularities are more evident in a limited set of high-frequency local transitions than in a unified whole-song sectional sequence.

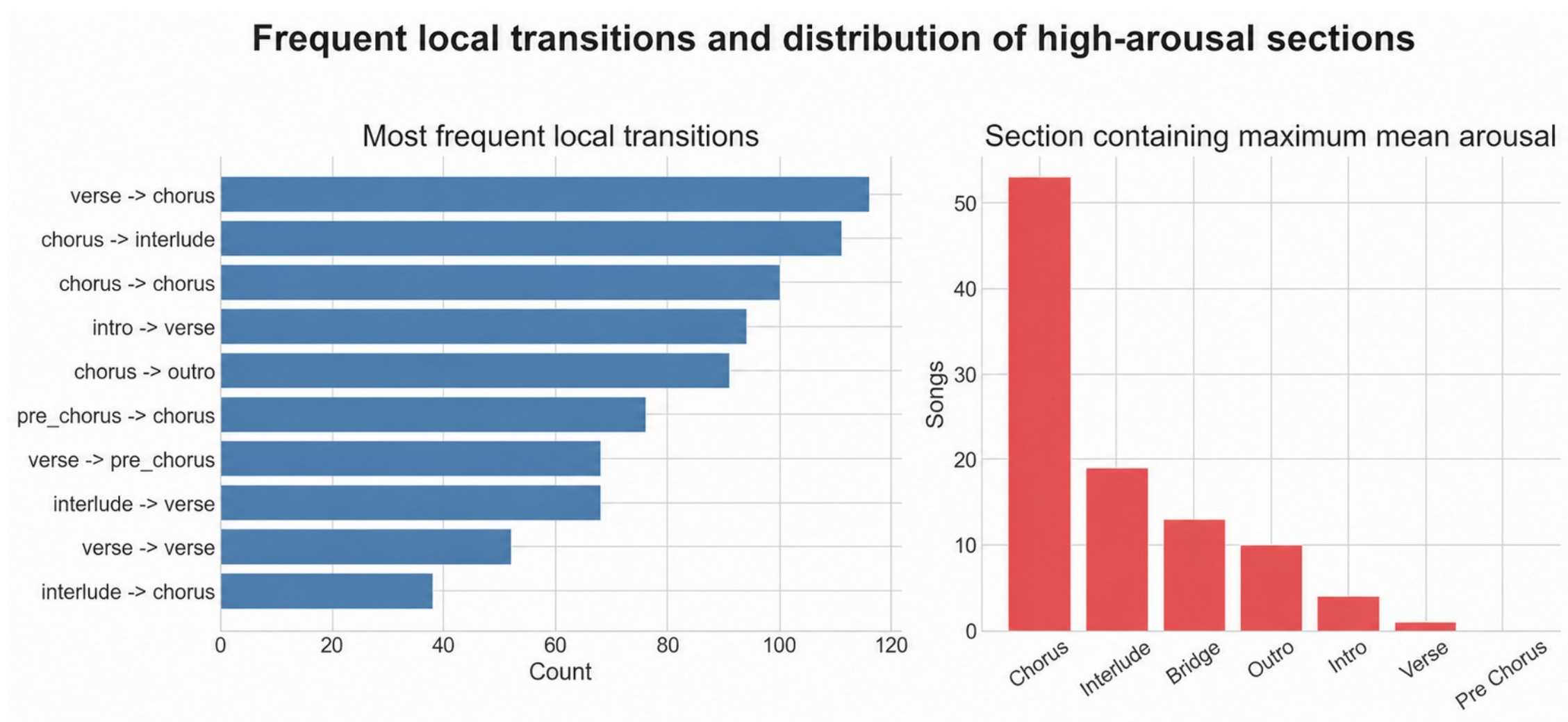

**Figure 2. High-Frequency Local Section Transitions and the Distribution of Section Types with the Highest Mean Arousal**

## 4.2 Emotional Functions of Sections: Chorus Primarily as an Arousal-Based Arrival

Song-weighted section means showed that Verse had the lowest mean Arousal (−0.073), compared with 0.206 for Chorus and 0.275 for Bridge. Within-song paired comparisons further reduced the influence of between-song baseline differences. Relative to Verse, Chorus showed a mean increase of 0.119 in Valence, with positive differences in 89% of songs, and a mean increase of 0.279 in Arousal, with positive differences in 99% of songs.

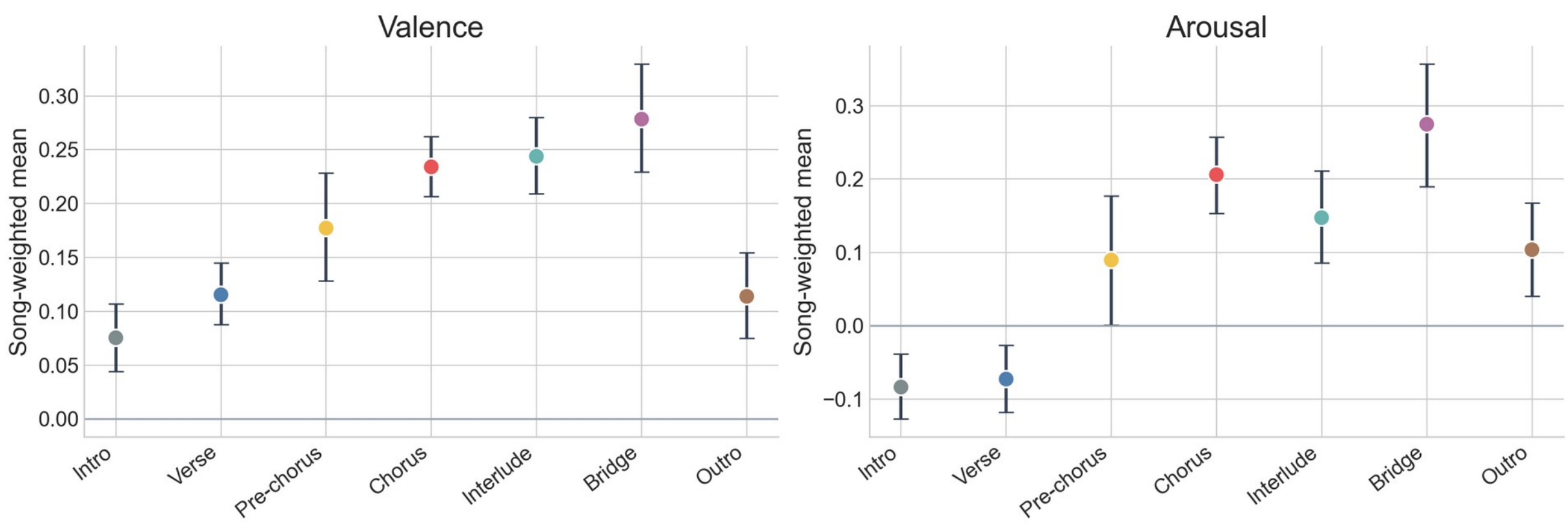

**Figure 3. Song-Weighted Mean Valence and Arousal across Seven Section Types with 95% Bootstrap Confidence Intervals**

**Table 2a. Within-Song Paired Comparisons of Valence**

| Valence Comparison | n | Mean Difference | 95% CI | Positice Proportion | Holm p |
|---|---|---|---|---|---|
| Chorus - Verse | 100 | +0.119 | [+0.100, +0.138] | 89.0% | <.001 |
| Pre-chorus - Verse | 38 | +0.055 | [+0.034, +0.076] | 76.3% | <.001 |
| Chorus - Pre-chorus | 38 | +0.078 | [+0.057, +0.099] | 92.1% | <.001 |
| Bridge - Chorus | 43 | +0.050 | [+0.019, +0.083] | 62.8% | .024 |
| Outro - Intro | 94 | +0.036 | [-0.001, +0.074] | 54.3% | .068 |

**Table 2b. Within-Song Paired Comparisons of Arousal**

| Valence Comparison | n | Mean Difference | 95% CI | Positice Proportion | Holm p |
|---|---|---|---|---|---|
| Chorus - Verse | 100 | +0.279 | [+0.251, +0.308] | 99.0% | <.001 |
| Pre-chorus - Verse | 38 | +0.135 | [+0.097, +0.172] | 84.2% | <.001 |
| Chorus - Pre-chorus | 38 | +0.200 | [+0.162, +0.243] | 100.0% | <.001 |
| Bridge - Chorus | 43 | +0.088 | [+0.040, +0.141] | 72.1% | .006 |
| Outro - Intro | 94 | +0.186 | [+0.131, +0.242] | 74.5% | <.001 |

Note for Tables 2a and 2b: Positive values indicate that the first section type has a higher value than the second. Statistical significance was assessed using two-sided Wilcoxon signed-rank tests.

Although Pre-chorus occurred in only 38 songs, it occupied a relatively consistent position in VA space when present. Compared with Verse, its mean Arousal was higher by 0.135, with positive differences in 84.2% of songs; Chorus was in turn higher than Pre-chorus by 0.200, with positive differences in all 38 songs. Given its structural position between Verse and Chorus, Pre-chorus therefore exhibited a relatively consistent buildup pattern in the present sample. Bridge showed higher Valence and Arousal than Chorus within the same songs by 0.050 and 0.088, respectively, indicating an overall tendency toward higher VA levels; however, this overall position does not imply that every transition into Bridge involved an increase. No clear Valence difference was detected between Outro and Intro, whereas model-estimated Arousal was higher in Outro by 0.186.

### 4.3 Transition Patterns: Diverse Modes of Arrival and Relatively Clear Paths of Closure

Strictly adjacent transitions showed that Arousal decreased by an average of 0.056 from Intro to Verse, indicating that many songs move from the Intro into a lower-arousal Verse, consistent with the Verse's role as a narrative baseline. Mean Arousal increased by 0.107 from Verse to Pre-chorus and by a further 0.154 from Pre-chorus to Chorus, indicating an overall upward progression across these two adjacent transitions in songs containing this structure.

**Table 3. Song-Weighted VA Changes across Key Adjacent Section Transitions**

| Adjacent Section Transition | No. of Events | ΔV | V 95% CI | ΔA | A 95% CI |
|---|---|---|---|---|---|
| intro->verse | 94 | -0.004 | [-0.020, +0.012] | -0.056 | [-0.082, -0.032] |
| verse->pre chorus | 68 | +0.037 | [+0.022, +0.052] | +0.107 | [+0.074, +0.138] |
| pre chorus->chorus | 76 | +0.062 | [+0.044, +0.079] | +0.154 | [+0.126, +0.187] |
| chorus->interlude | 111 | +0.034 | [+0.017, +0.052] | +0.010 | [-0.011, +0.032] |
| interlude->verse | 68 | -0.018 | [-0.037, +0.001] | -0.048 | [-0.080, -0.017] |
| chorus->bridge | 33 | +0.008 | [-0.021, +0.038] | +0.030 | [-0.008, +0.070] |
| bridge->chorus | 31 | -0.006 | [-0.037, +0.023] | +0.024 | [-0.003, +0.052] |
| chorus->outro | 91 | -0.132 | [-0.159, -0.105] | -0.188 | [-0.230, -0.147] |

Note: When the same transition occurred multiple times within a song, the transition values were first averaged within that song before cross-song bootstrap analysis.

Transitions into Chorus did not follow a single VA pattern. Among all 362 transitions whose target section was Chorus, using ±0.05 as a descriptive threshold, both Valence and Arousal increased in 138 cases (38.1%), Arousal alone increased markedly in 71 cases (19.6%), and both dimensions remained approximately stable in 64 cases (17.7%). A smaller number of transitions showed other patterns, including decreased Valence with increased Arousal or decreases in both dimensions. Overall, Chorus tended to occupy a high-Arousal position, but songs reached it through multiple VA pathways, including joint increases in Valence and Arousal, predominantly Arousal-based increases, and high-level continuation.

The Arousal change from Chorus to Interlude was small (ΔA=0.010, with the 95% CI including zero), while Valence increased slightly by 0.034. By contrast, Interlude→Verse showed a clearer decrease in Arousal (ΔA=−0.048), suggesting that Interlude may more often function as a transitional or energy-maintaining section in the present sample rather than producing an immediate decline upon entry. Transitions into and out of Bridge did not show a consistent direction of Arousal change, indicating greater variability in its local transition patterns. Chorus→Outro showed decreases of 0.132 in Valence and 0.188 in Arousal, representing a relatively consistent local closing pattern among the high-frequency transitions examined.

## 4.4 Repeated Sections: VA Differences between First and Last Occurrences

For section types occurring at least twice within a song, comparisons between the first and last occurrences showed significant increases in both Valence and Arousal for Verse, Pre-chorus, Chorus, and Interlude, with the corresponding singer-clustered bootstrap intervals remaining positive. From the first to the last occurrence, Chorus showed an average Arousal increase of 0.225, with increases observed in 90% of songs, while Interlude showed an increase of 0.260, with increases in 93.9% of songs. These results suggest that repeated section types did not simply return to their initial VA levels at their final occurrence, but often appeared at higher VA states.

**Table 4. VA Differences between the First and Last Occurrences of Repeated Sections**

| Section Type | n | First-to-Last ΔV | Positive V | First-to-Last ΔA | Positive A | Maximum Holm p |
|---|---|---|---|---|---|---|
| Verse | 91 | +0.099 | 81.3% | +0.165 | 82.4% | <.001 |
| Pre-chorus | 37 | +0.112 | 91.9% | +0.156 | 91.9% | <.001 |
| Chorus | 100 | +0.085 | 71.0% | +0.225 | 90.0% | <.001 |
| Interlude | 33 | +0.133 | 84.8% | +0.260 | 93.9% | <.001 |

Note: The maximum Holm-adjusted p value refers to the larger of the adjusted p values for Valence and Arousal within each section type; both dimensions were significant for all four section types.

The first-to-last Arousal difference for Chorus was further compared with the corresponding differences for Verse, Pre-chorus, and Interlude within the same songs. The three Holm-adjusted *p* values were 0.142, 0.214, and 0.751, respectively, none of which reached statistical significance. Thus, the present data do not support the claim that the first-to-last Arousal change in Chorus was significantly greater than that observed for other repeated section types. Higher VA values at the final occurrence were therefore not unique to Chorus, as similar tendencies were also observed for Verse, Pre-chorus, and Interlude. Accordingly, the later increase in VA should not be attributed solely to the final Chorus.

## 4.5 Whole-Song Trajectories: Local Directional Changes and Later Arousal Elevation

Across the 100 pop songs, the mean VA path length was 1.443, whereas the mean straight-line displacement from the first to the last section was 0.316, yielding a mean path directness of 0.242. Overall, the VA trajectories were not characterized by a simple unidirectional increase, but instead involved multiple local reversals and directional changes across sections.

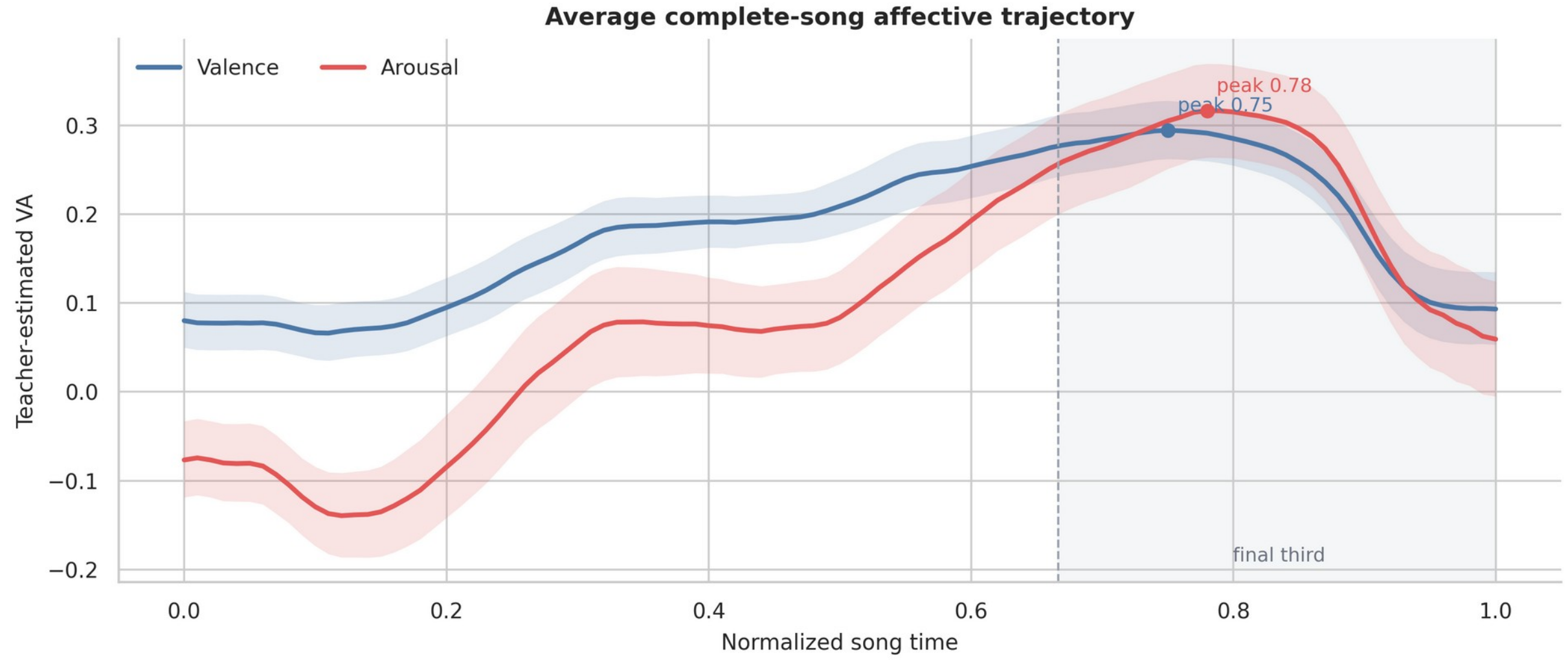

**Figure 4. Mean VA Trajectories across 100 Songs after Normalization to 101 Time Points, with 95% Song-Level Bootstrap Confidence Bands (Gray Area Indicates the Final Third of the Song)**

Despite this local variability, the average trajectories still showed a tendency toward higher values in the later part of the song. From the first to the last section, the mean net change in Valence was +0.039, with positive changes in 56% of songs, whereas the mean net change in Arousal was +0.193, with positive changes in 75% of songs. The peaks of the average Valence and Arousal trajectories occurred at normalized time positions of 0.75 and 0.78, respectively. Across individual songs, the mean and median normalized positions of the section with the highest mean Arousal were 0.733 and 0.755, and 79 songs placed this section within the final third of the song. This indicates that high-arousal sections occurred more frequently in the later part of the song.

In 53 songs, the section with the highest mean Arousal was a Chorus; the corresponding section was an Interlude in 19 songs, a Bridge in 13, an Outro in 10, an Intro in 4, and a Verse in 1. Thus, the section type associated with the highest Arousal was not fixed. The final Chorus was the section with the highest mean Arousal in only 31 songs. In addition, the largest VA-distance transition between adjacent sections occurred upon entry into Outro in 44 songs and into Chorus in 27 songs, with the remaining cases distributed across Verse, Pre-chorus, Interlude, and Bridge. These results suggest that substantial VA changes in the later part of a song are associated not only with high-arousal sections but also frequently with transitions into Outro.

### 4.6 Robustness Analysis

After singer-clustered bootstrap resampling across the 60 quota artists, the directions of the core comparisons were broadly consistent with those obtained from the song-level analyses, except that the Valence difference between Outro and Intro remained inconclusive. This suggests that the main findings were not strongly driven by a small number of artists contributing multiple songs. In addition, structural and duration-related measures did not show clear differences across ranking tiers after correction for multiple comparisons. Because chart ranking was used only for sampling stratification, no further interpretation was made regarding its relationship with song form or emotional characteristics.

## 5 Discussion

### 5.1 From Emotional Polarity to Arousal-Dominated Formal Dynamics

In the present study, the more stable emotional differences across sections were reflected primarily in Arousal rather than Valence. This pattern may be related to the ways in which structural reinforcement is realized in popular songs. Formal progression from Verse through Pre-chorus to Chorus is often accompanied by changes in vocal intensity, textural density, register, rhythmic drive, and production layers. These parameters primarily alter the energy and activation level of the music and may therefore map more directly onto Arousal. By contrast, the emotional polarity represented by Valence may be jointly influenced by factors such as lyrical semantics, harmonic progression, and specific musical context, whose directions of change vary across songs and are therefore less likely to produce consistent sectional differences. Accordingly, the more stable characteristic of Chorus relative to Verse is not that it is "more positive," but that it exhibits a higher level of activation; the Verse-Pre-chorus-Chorus progression may likewise be understood as a process of formal reinforcement realized primarily through changes in energy and density. This interpretation remains a mechanistic hypothesis in the present study and requires further validation using specific acoustic and lyrical features.

### 5.2 Local Transition Regularities and Diversity in Whole-Song Form

The 76 distinct raw sequences and 49 collapsed sequences indicate that the selected songs did not conform to a single whole-song sectional template. At the same time, the eight most frequent transition types accounted for 76.5% of all adjacent transitions, suggesting that cross-song commonalities were more evident in local sectional relationships. Paths such as Verse→Chorus, Verse→Pre-chorus→Chorus, Chorus→Interlude→Verse, and Chorus→Outro recurred across the sample, yet the VA changes associated with entering or leaving the same section type were not uniform. For example, Chorus could be reached through gradual buildup, a more direct increase, or continuation at an already elevated VA level. This suggests that section labels are better understood as functional contexts for interpreting emotional change than as fixed rules determining the direction of VA variation. Overall, relatively concentrated local transitions and diverse whole-song configurations can coexist, allowing pop songs to retain a degree of structural recognizability while accommodating multiple organizational patterns, including consecutive Choruses, Interludes, and Bridges occurring at different positions.

### 5.3 Formal Recurrence and Changes in Emotional State

Comparisons between the first and last occurrences of repeated sections showed that the VA state associated with a given section type did not necessarily return to its initial level when the section recurred. The final occurrences of Verse, Pre-chorus, Chorus, and Interlude all exhibited higher Valence and Arousal, with the Arousal differences showing particularly consistent patterns. This suggests that formal recurrence in pop songs is often accompanied by changes in emotional state, with the same sectional function being re-realized within a different whole-song context. Notably, the first-to-last Arousal difference for Chorus was not significantly greater than the corresponding differences for Verse, Pre-chorus, and Interlude, indicating that later-stage increases in VA are not unique to Chorus but more likely reflect a common change in the song's overall emotional progression. Although the same functional sections recur formally, their emotional states change as the song unfolds, giving rise to a pattern of "formal recurrence without complete emotional recurrence." At the whole-song level, this change does not take the form of a sustained monotonic increase; rather, through local pullbacks, shifts, and renewed arrivals, later occurrences of the same section type tend to occupy higher VA states. Together with the tendency for high-Arousal positions to occur more frequently in the later part of a song, this constitutes a non-monotonic process of accumulation. The present study therefore summarizes this organizational pattern as "local cycles−global accumulation."

### 5.4 Climax Formation and Climax Release

A stronger final Chorus is not equivalent to the final Chorus being the song's climax. Although Arousal was often higher in the final occurrence of Chorus than in its first occurrence, the final Chorus was also the section with the highest mean Arousal in only 31 songs. High-Arousal positions could also occur in Interlude, Bridge, and Outro. At the same time, the largest VA-distance transition between adjacent sections occurred upon entry into Outro in 44 songs, and Chorus→Outro showed a relatively consistent simultaneous decrease in Valence and Arousal. From this perspective, the establishment of a high-arousal state and its subsequent release and closure can be viewed as complementary components of emotional organization in the later part of a song, with Outro potentially serving a distinct closing function. Key emotional events in a song concern not only how a climax is reached, but also how it is exited.

### 5.5 Implications for MER and Pop Music Analysis

For MER, a single whole-song emotion label cannot adequately preserve the dynamic information associated with sectional changes within a song. Integrating continuous VA trajectories with sectional structure makes it possible to examine emotional variation at different formal positions, shifting the analytical focus from “what emotion does this song express?” to “how does emotion unfold through the song’s structure?” This perspective may inform future work on emotion-based retrieval, structure-controllable music generation, and evaluation of generated music, although its practical utility should be further examined using larger datasets and task-specific evaluations.

For pop music analysis, the present findings show several points of correspondence with established theories of musical form. Summach (2011) proposed that Pre-chorus functions to propel the subsequent arrival of Chorus; the progressive Arousal pattern observed here provides a quantitative complement to this formal interpretation. At the same time, the variability in transitions involving Bridge and in the VA pathways leading into Chorus is consistent with de Clercq’s (2017) view that formal labels are context-dependent rather than acoustically uniform. Section types are therefore better understood as functional contexts for interpreting emotional change than as fixed emotional or acoustic templates. Future research could further integrate features such as loudness, rhythm, register, harmony, timbre, vocal expression, and lyrical content to directly test the Arousal-dominated mechanism of structural reinforcement proposed above, and to compare how similar sectional functions are realized across different songs.

## 5.6 Scope of the Study and Future Work

The conclusions of this study are based primarily on the selected sample of 100 Chinese pop songs and on VA trajectories generated by frozen teacher models, and should therefore be interpreted as relative trends within this sample. Section annotations were primarily conducted by the researcher, with 10 songs additionally reviewed by a musically trained reviewer. As systematic inter-rater agreement was not evaluated across the full corpus, some functional boundaries may retain a degree of interpretive flexibility. Differences between first and last occurrences primarily describe changes associated with earlier versus later positions in the song and should not be interpreted as evidence that repetition itself causes strengthening.

Future work may extend the corpus across a broader range of periods, styles, platforms, and artists, and incorporate additional human continuous-VA annotations to compare sectional emotion patterns across different datasets. Acoustic and performance-related features, including loudness, rhythm, register, harmony, timbre, lyrics, and vocal characteristics, could also be integrated to investigate how similar emotional functions are realized through different musical means. From an application perspective, section-aware dynamic emotion analysis may further be explored in music retrieval, recommendation, structure-controllable generation, and evaluation of generated music.

## 6 Conclusion

This study examined 100 Chinese pop songs by integrating 1,046 manually annotated functional sections with continuous VA trajectories and investigating the relationship between sectional structure and emotional dynamics from the perspectives of section type, adjacent transitions, repeated sections, and whole-song trajectories. The results indicate that emotional differences across sections were reflected more strongly in Arousal. Verse typically formed a relatively low-activation baseline, Pre-chorus showed a progressive buildup when present, and Chorus served as a more pronounced high-arousal point of arrival, whereas Interlude, Bridge, and Outro tended to exhibit transitional, contextually divergent, and closing functions, respectively. At the same time, VA pathways into the same section type were not uniform, suggesting that section labels are better understood as functional contexts for interpreting emotional change than as fixed emotional templates.

At the whole-song level, the sample showed diverse sectional configurations alongside relatively concentrated local transition patterns. The same section type could occupy different VA states at earlier and later positions in a song, and whole-song trajectories were not characterized by simple unidirectional progression. Instead, they involved local buildup, reversal, and renewed arrival, together with a tendency toward elevated Arousal in the later part of the song. High-arousal positions occurred more frequently toward the end of songs but were not fixed to the final Chorus. In addition, the relatively large VA changes observed upon entry into Outro, together with the simultaneous decline in Valence and Arousal from Chorus to Outro, suggest that release and closure following high-arousal states also constitute an important part of later-stage emotional organization.

On this basis, the emotional organization of the selected Chinese pop songs is summarized as an empirical framework of “local cycles-global accumulation.” Rather than sharing a single whole-song template, the songs are characterized by relatively concentrated local transitions combined with diverse whole-song configurations. Within this process, local sections may recur, and emotional trajectories may undergo pullbacks and renewed arrivals; however, later occurrences of the same section type tend to occupy higher VA states, while high-Arousal positions are more frequently distributed in the later part of the song. Thus, “global accumulation” refers to a later-stage upward tendency that coexists with local fluctuations, rather than to a sustained monotonic increase in VA over the course of a song. The overall pattern can be further summarized as local cycling, later-stage accumulation, delayed climax, and release-based closure. This framework provides a section-aware perspective for linking continuous music emotion recognition with popular-music form analysis and offers a basis for future research on acoustic mechanisms, cross-corpus comparison, and structure-controllable music generation.